\documentclass[twocolumn,aps,prb,superscriptaddress,a4paper,nofootinbib,showpacs,showkeys]{revtex4-1}

\setcitestyle{numbers,square}

\usepackage{graphicx}
\usepackage{type1cm}
\usepackage[normalem]{ulem}  
\usepackage{bm,dcolumn,amssymb,amsmath,braket,siunitx,color}
\usepackage{sidecap}

\PassOptionsToPackage{hyphens}{url}

\newcommand{\mm}[1]{\mbox{$#1$}}
\newcommand{\dstd}{\mathrm{d}}

\newcommand{\bsi}{\mbox{$\beta$-Si$_{6-z}$Al$_{z}$O$_{z}$N$_{8-z}$}}
\renewcommand{\sin}{\mbox{Si$_{3}$N$_{4}$}}

\newcommand{\betasin}{\mbox{$\beta$-Si$_{3}$N$_{4}$}}
\newcommand{\wtk}{\mbox{\sc wien2k}}
\newcommand{\kkr}{\mbox{\sc sprkkr}}

\newcommand{\eg}{\mbox{$E_{g}$}}

\newcommand{\two}{1$\times$1$\times$2}
\newcommand{\three}{1$\times$1$\times$3}
\newcommand{\bcb}{the bottom of the conduction band}
\newcommand{\tvb}{the top of the valence band}
\newcommand{\ndiff}{\mbox{$n_{\text{Si}_{6-z}\text{Al}_{z}\text{O}_{z}\text{N}_{8-z}}-n_{\text{Si}_{3}\text{N}_{4}}$}}  
\newcommand{\ea}{{\it et al.}}

\begin{document}

\title{Dependence of the electronic structure of
  \mbox{$\beta$-Si$_{6-z}$Al$_{z}$O$_{z}$N$_{8-z}$} on the (Al,O)
  concentration~$z$ and on the temperature}



\author{Saleem \surname{Ayaz Khan}} \affiliation{New Technologies
  Research Centre, University of West Bohemia, CZ-301~00~Pilsen, Czech
  Republic}

\author{Ond\v{r}ej \surname{\v{S}ipr}} \affiliation{New Technologies
  Research Centre, University of West Bohemia, CZ-301~00~Pilsen, Czech
  Republic} \affiliation{FZU -- Institute
  of Physics of the Czech Academy of Sciences, Cukrovarnick\'{a}~10,
  CZ-162~53~Prague, Czech Republic }

\author{Ji\v{r}\'{\i} \surname{Vack\'{a}\v{r}}} \affiliation{FZU --
  Institute of Physics of the Czech Academy of Sciences, Na
  Slovance~2, CZ-182~21~Prague, Czech Republic }

\author{J\'{a}n \surname{Min\'{a}r}} \affiliation{New Technologies
  Research Centre, University of West Bohemia, CZ-301~00~Pilsen, Czech
  Republic}

\date{\today}

\begin{abstract}
  \bsi\ is a prominent example of systems suitable as hosts for
  creating materials for light-emitting diodes (LEDs).  In this work,
  the electronic structure of a series of semiordered and disordered
  \bsi\ systems is investigated by means of {\em ab initio}
  calculations, using the FLAPW and Green function KKR methods.
  Finite temperature effects are included by averaging over
  thermodynamic configurations within the alloy analogy model.  We
  found that the dependence of the electronic structure on the (Al,O)
  concentration~$z$ is similar for semiordered and disordered
  structures.  The electronic band gap decreases with increasing~$z$
  by about 1.5~eV when going from~$z$=0 to~$z$=2.  States at the top
  of the valence band are mostly associated with N atoms whereas the
  states at the bottom of the conduction band are mostly derived from
  O~atoms.  Increasing the temperature leads to a shift of the bottom
  of the conduction band to lower energies.  The amount of this shift
  increases with increasing~$z$.
\end{abstract}

\maketitle


\section{Introduction}   \label{sec:intro}

\bsi\ doped with rare earth ions is one of the intensively studied
classes of materials suitable for light-emitting diodes (LEDs).  In
particular, Eu-doped \bsi\ with $z$=0.17 is a promising green phosphor
candidate for creating white light in phosphor-converted LEDs
\cite{HXK+05}.  One of its attractive properties is a good thermal
stability \cite{XHL+07,RPW+08,YEI+12}.  The \bsi:Eu$^{2+}$ system can
be prepared in many modifications --- one can vary the (Al,O)
concentration~$z$ as well as of the Eu concentration.  A lot of
studies have been devoted to it and a lot of results have been
gathered.  However, also due complexity of the system, the conclusions
are sometimes controversial and the knowledge scattered.  For example,
the ranges of~$z$ employed in different studies often do not overlap,
so a direct comparison is difficult.

The first step in studying rare-earth-doped \bsi\ is investigating the
host system, which derives from hexagonal $\beta$-\sin.  A lot of
attention was paid to the structure of \bsi.  The studies performed up
to now indicate that the structure is neither crystalline nor totally
random; probably there is some degree of ordering
\cite{DMH+87,LRS+96,OI+01,SHK+01,TTI+02,FM+03,WYC+16,CGL+17,ZFT+17,SKM+20}.
If we 
describe the structure of parental $\beta$-\sin\ via the space group
176 ($P6_{3}/m$), the Si atoms are all equivalent occupying 6$h$
Wyckoff positions, whereas the N atoms are split into two groups,
occupying 6$h$ positions and 2$c$ positions.  In \bsi, Al
atoms substitute Si atoms and O atoms substitute N atoms.  One of the
questions is whether O atoms go to 6$h$ or to 2$c$ positions.  Some
experiments suggest a slight preference of O atoms for the
6$h$ sites \cite{DMH+87,LRS+96,SHK+01} but the results are not
conclusive.  Some studies do not indicate any preference
for the positions of the O atoms \cite{TTI+02,BHL+03}.  Recent 
theoretical simulations of Wang \ea\ \cite{WYC+16} indicate that O
atoms should substitute N atoms at the 2$c$ positions.  Increase of
the (Al,O) concentration~$z$ leads to less ordering \cite{ZFT+17}.

An important characteristics of all luminescent materials is the
electronic band gap \eg. There is common agreement that the gap of
\bsi\ decreases with increasing~$z$.  However, conflicting results
have been obtained about the rate of this decrease.  Calculations of
Ching \ea\ \cite{CHM+00} and Boyko \ea\ \cite{BGS+14} suggested a very
rapid decrease of \eg\ with increasing~$z$.  On the other hand,
Hirosaki \ea\ \cite{HKO+05} and Wang \ea\ \cite{WYC+16} found a much
slower decrease. 

Linked to this is the question about the mechanism which leads to the
decrease of the gap with increasing~$z$, and about the character of
the states at the top of the valence band and at the bottom of the
conduction band.  Calculations of Boyko \ea\ \cite{BGS+14} and Wang
\ea\ \cite{WLQ+19} suggest that the bottom of the conduction band is
formed by impurity-like states related to O atoms.  Other calculations
do not confirm this \cite{HKO+05,WYC+16} and indicate that the bottom
of the conduction band is composed mainly by states associated
  with Si and N atoms \cite{WYC+16}.

Another interesting issue is the thermal quenching of the
luminescence.  The fact that the thermal stability of Eu-doped \bsi\
is very good seems to be well established \cite{XHL+07,RPW+08}.
However, there have been conflicting reports how this stability varies
upon changes of~$z$.

It follows that the understanding of \bsi\ is still patchy and some
issues are unclear or even controversial.  In particular this concerns
the dependence of the gap \eg\ on the (Al,O) concentration and the
character of the electronic states at the bottom of the conduction
band.  Also it would be interesting to learn which states are most
affected if the temperature is varied and how this depends on the (Al,O)
concentration $z$.

Because different studies led to different outcomes, there
is a need for a comprehensive study which would employ more
calculational approaches, so that robust conclusions can be drawn.
The fact that the structure of \bsi\ is neither fully ordered nor
disordered should be taken into account.  The range of~$z$ for the
\bsi\ systems should be such that is covers both large values
($z\approx2$), where the effect of introducing (Al,O) atoms will be
most visible, and small values ($z\approx0.1$--0.3), so that the
composition will be similar to the composition of technologically
interesting materials.

Therefore, we present in this paper a theoretical study of the
electronic structure for a series of \bsi\ systems, performed by
employing (i) the full-potential augmented plane waves (FLAPW) method
as implemented in the \wtk\ code and (ii) the Green function
multiple-scattering or Korringa-Kohn-Rostoker (KKR) method as
implemented in the \kkr\ code.  We investigate semiordered systems
(described by means of supercells) as well as disordered systems,
treated within the coherent potential approximation (CPA).  Finite
temperature effects are accounted for by employing the alloy analogy
model to perform averaging over configurations in agreement with their
thermodynamic weights.

Our focus is not on attempting to find the most energetically
  stable geometries but rather on how the local variations in the
  geometric structure and the degree of disorder can influence the
  electronic structure of \bsi, particularly concerning the 
  states around the gap.  Given the uncertainties concerning the
  structure of \bsi, it will be useful to find, e.g., what impact it
  can have if the sites preference of O atoms changes from $2c$ to
  $6h$, or which aspects of the electronic structure are robust or
  volatile with respect to disorder and which states will be most
  affected if the temperature increases.


\section{Methodological framework}


\subsection{Systems}    \label{sec:systems}

\begin{figure}
\includegraphics[width=8.4cm]{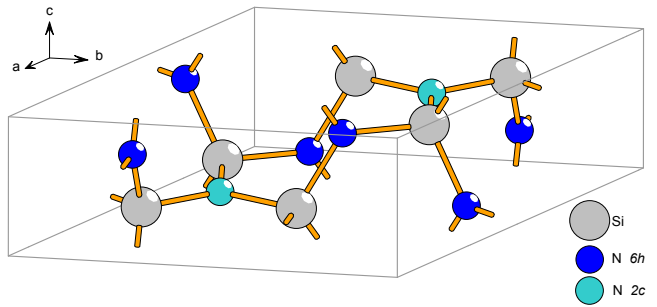}%
\caption{Perspective diagram of the unit cell of
  $\beta$-\sin. Two inequivalent N sites are distinguished, as
  indicated in the legend.}
\label{fig:struct}
\end{figure}

We investigate \bsi\ systems derived from the parental $\beta$-\sin\
compound by substituting Si with Al and N with O.  The structure of
$\beta$-\sin\ is hexagonal, described by the space group 176
($P6_{3}/m$), with lattice constants $a$=7.655~\AA\ and $c$=2.928~\AA.
The unit cell contains two formula units. All Si atoms are equivalent,
located at $6h$ Wyckoff positions. The N atoms split into two groups,
located at $2c$ positions (two sites) and at $6h$ positions (six
sites).
The two N sites differ only slightly concerning their nearest
  neighborhood.  In both cases, N atoms are threefold coordinated.
  The Si--N bond lengths are 1.742~\AA\ for the $2c$ site and
  1.747~\AA\ for the $6h$ site.  The bond angles are $120^{\circ}$ for
  the $2c$ site and $123^{\circ}$ and $114^{\circ}$ for the $6h$ site.
  A structural diagram is shown in
  Figure~\protect\\ref{fig:struct}.

If \betasin\ is doped to form \bsi, Al atoms
substitute Si atoms at $6h$ positions whereas for O atoms, it is
unclear whether they preferentially substitute N atoms at $2c$
positions or N atoms at $6h$ positions
\cite{LRS+96,CHM+00,HKO+05,BGS+14,WYC+16}.  It is possible that doping
\bsi\ further with rare earths will affect the preference of O being
at either location.  As one of our motivations is to study \bsi\ as a
host material for \bsi:Eu$^{2+}$, we will deal with two types of
configurations: where O atoms prefer $2c$ positions and where they
prefer $6h$ positions, to see the difference.

Earlier studies indicate that the distribution of (Al,O) atoms in
the parental $\beta$-\sin\ lattice is neither fully ordered nor
completely disordered \cite{HKO+05,WYC+16}.  Studying the influence of
disorder is one of our goals.  Therefore, we split the systems into
two classes, semiordered and disordered, keeping in mind that
properties of real samples will be somewhere between properties of
semiordered systems and properties of disordered system.

First, we introduce semiordered systems.  By that we mean that for
each (Al,O) concentration $z$, we take several geometric configurations
with a different distribution of (Al,O) atoms among the available
positions and when calculating properties such as the density of
states (DOS), we make a simple arithmetic average over all
  such configurations.  To describe the system for each 
$z$, we select six configurations built by means of supercells of the
parental $\beta$-\sin.  We use \three\ supercells for the~$z$=0.333
composition and \two\ supercells for the~$z$=0.5, $z$=1, and $z$=2
compositions.  The positions of the (Al,O) atoms were selected {\em
  ad hoc} for each supercell, requiring only that an Al atom is always
sitting next to a O atom \cite{OI+01,WYC+16,CGL+17}.  We further distinguish
systems where O atoms are put into $2c$ positions and where O atoms
are put into $6h$ positions. Each supercell was structurally relaxed.
The crystallographic data for each of the relaxed supercell have been
deposited at the CCDC/FIZ Karlsruhe crystal structures deposition
service under deposition numbers 2172561--2172621
\cite{ccdcdef}. 

For disordered systems we assume a complete substitutional disorder.
This means that sites occupied in the parental \betasin\ by Si are now
occupied partially by Si and partially by Al (with appropriate
concentrations) and, similarly, $2c$ or $6h$ positions occupied
originally by N are now occupied partially by N and partially by O,
depending on whether we want to deal with system where the O atoms
prefer $2c$ positions or where they prefer $6h$ positions.  The
geometric  positions of the sites  are the same as for the original 
  \betasin\ structure.


\subsection{Computational methods}    \label{sec:method}

The calculations were performed within the ab-initio framework of the
density functional theory.  Structural relaxations of configurations
which represent semiordered systems were performed by means of the
plane waves pseudopotential method using the {\sc vasp} code
\cite{KF+96}.  The electronic structure of the relaxed supercells was
calculated by means of the FLAPW method implemented in the {\sc
  wien2k} code \cite{Blaha+01}.  Systems with substitutional disorder
were treated by means of the Green-function multiple-scattering
Korringa-Kohn-Rostoker (KKR) method using the the {\sc sprkkr} code
\cite{sprkkr-code}.  The disorder was treated within the coherent
potential approximation (CPA) \cite{EKM11}.  Finite temperature
effects were included by the so-called alloy analogy model
\cite{EMC+15}: temperature-induced atomic displacements were treated
as localized and uncorrelated, giving rise to an additional type of
disorder that can be described using the CPA.
For further technical details we refer to the Appendix~\ref{sec:compdets}.

When relaxing the geometries of the supercells representing the
semiordered systems, we proceeded in two steps.  First, the external
parameters, i.e., the lattice constants $a$ and $c$ were optimized,
keeping the positions of atoms in the supercell same as for undoped
$\beta$-Si$_3$N$_4$. Once the external parameters have been found, an
additional relaxation of the the internal parameters, i.e., of the
atomic positions of individual atoms, was performed.  We do this to
account for the fact that a real \bsi\ sample will contain many
supercells simultaneously and the lattice parameters will be
determined also by interaction with other subunits where the local
structure may be different than in the supercell we select.  The
two-step procedure allows us to perform local structure relaxation
while limiting at the same time the danger that the long-range
structure might be excessively affected by nonrepresentative local
effects.


\section{Results and discussion}   \label{sec:res}


\subsection{Dependence of lattice parameters on~$z$}

\label{sec:lattice}

\begin{figure}
\includegraphics[width=8.4cm]{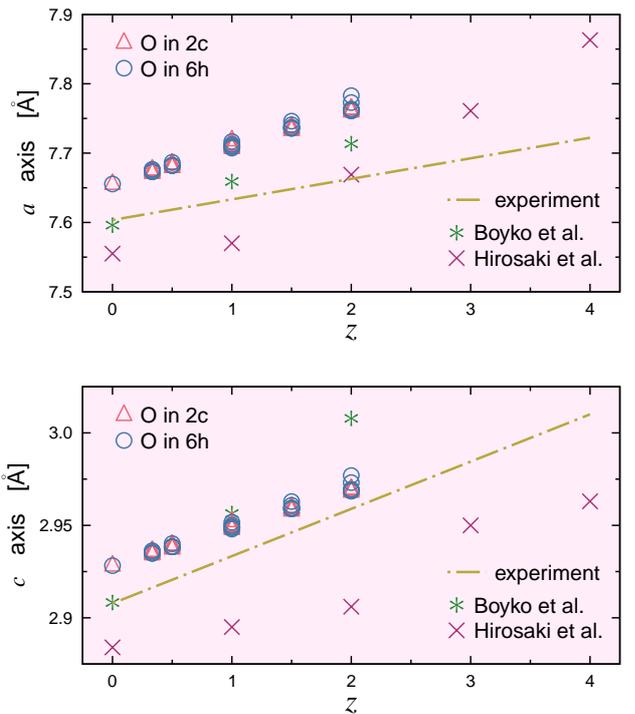}%
\caption{Dependence of the $a$ axis (top graph) and the
  $c$ axis (bottom graph) of \bsi\ on~$z$. For each $z$, the system is
  represented by six supercells, with O atoms either at $2c$ positions
  (red triangles) or at $6h$ positions (blue circles).  Results of
  earlier calculations of Hirosaki \ea\ \cite{HKO+05} and of Boyko
  \ea\ \cite{BGS+14} are shown for comparison.  Fits to the
  experimental data of Ekstr\"{o}m \ea\ \cite{EKN+89} are shown by
  straight lines.}
\label{fig:cell}
\end{figure}

We performed structural relaxation for
each representing supercell.  The results
for the dependence of the lattice constants $a$, $c$ on the (Al,O)
concentration~$z$ are presented in Figure~\ref{fig:cell}. The markers
representing the data for individual supercells nearly coincide for
each~$z$. This means that the lattice constants do not strongly depend
on the specific positions of (Al,O) atoms in the supercell.  Moreover,
the
trends do not depend on whether the~O atoms are at the 2$c$ or at the
6$h$ positions.  Comparing with other studies, one sees that the
experimentally observed increase of $a$ and $c$ with increasing~$z$ is
fairly reproduced by any calculation \cite{BGS+14,HKO+05,EKN+89}.

In the following, whenever we investigate semiordered structures, we
deal with averaging of results obtained for the relaxed supercells.
When investigating disordered structures, we keep the atoms at the
original $2c$ or $6h$ positions but the lattice constants $a$ and $c$
are taken those obtained for semiordered structures, as given in
Figure~\ref{fig:cell}.


\subsection{Band gap}

\label{sec:gap}

\begin{figure}
\includegraphics[width=8.4cm]{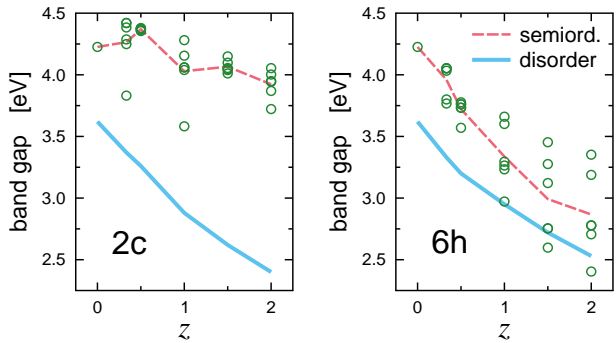}%
\caption{Dependence of the energy band gap~$E_{g}$
  on~$z$ for semiordered and disordered structures.  The left graph is
  for O atoms at 2$c$ positions, the right graph is for O atoms at
  6$h$ positions.  Values for individual supercells are shown via
  circles, values obtained by averaging over individual supercells 
  are shown by dashed lines.  Data obtained for
  disorded systems via the CPA are shown via solid lines.}
\label{fig:gap}
\end{figure}

The dependence of the energy band gap $E_{g}$ on~$z$ is depicted in
Figure~\ref{fig:gap}.
It was evaluated based on the energy difference between the
  states at \tvb\ and at \bcb\ and serves as an approximation for the
  optical gap.
For semiordered systems described by the
supercells, $E_{g}$ depends strongly on the particular positions of Al
and O atoms in the supercell: there is a {\em large spread} of values
of $E_{g}$ around the average.  As a whole, the gap $E_{g}$
significantly depends on the position of O atoms.  If O atoms are at
$2c$ positions, the decrease of $E_{g}$ with~$z$ is slower than if O
atoms are at $6h$ positions (cf.\ the dashed lines in the left graph
and in the right graph of Figure~\ref{fig:gap}).

The large scatter of
\eg\ values for supercells corresponding to the same~$z$ suggests that
preparation conditions might crucially affect the properties of
specific \bsi\ and also of \bsi:Eu$^{2+}$ samples.  Namely, different
preparation methods may lead to different distributions of (Al,O)
atoms among the Si and N sites which in turn will lead to different
electronic properties, as evidenced by Figure~\ref{fig:gap}. 

The situation is different for fully disordered systems treated within
the CPA, where the gap $E_{g}$ is practically the same no matter
whether the O atoms are at $2c$ or at $6h$ positions.  It should be
noted that the change of~$E_{g}$ with~$z$ is mostly due to {\em
  changes in the chemical composition} and not due to changes in the
lattice parameters.  We checked that if the calculations are done with
fixed lattice constant (same~$a$ and~$c$ for all~$z$'s), the trends
of~$E_{g}$ with~$z$ are very similar as if the lattice constant is
varied.

The decrease of $E_{g}$ with increasing~$z$ demonstrated by
Figure~\ref{fig:gap} is similar to what can be estimated from earlier
calculations of Hirosaki \ea\ \cite{HKO+05} and Wang \ea\
\cite{WYC+16}.  On the other hand, calculations of Ching \ea\
\cite{CHM+00} and of Boyko \ea\ \cite{BGS+14} indicated a much quicker
decrease of \eg\ with increasing~$z$.  Specifically, Ching \ea\
\cite{CHM+00} observed a decrease from about 4~eV for~$z$=0 to about
1.3~eV for~$z$=1 and Boyko \ea\ \cite{BGS+14} observed a decrease from
4.3~eV for~$z$=0 to 1.8~eV or 2.2~eV for~$z$=2 (depending on the
structural model).  The difference between the results of Hirosaki
\ea\ \cite{HKO+05}, Wang \ea\ \cite{WYC+16} and ours on the one hand
and of Ching \ea\ \cite{CHM+00} and Boyko \ea\ \cite{BGS+14} on the
other hand is quite big.  The reason is unclear.  The trend of the
experimental data is similar to our results: there is a decrease of
\eg\ from 7.2~eV for~$z$=0 to 6.2~eV for~$z$=2 \cite{BGS+14}.  The
absolute values of \eg\ provided by the calculations are smaller than
values obtained from experiment, which is a well-known problem of the
GGA.


\subsection{Dependence of DOS on $z$}

\label{sec:dos}

An overall view on how the DOS changes upon varying the (Al,O)
concentration~$z$ is presented in the Appendix~\ref{sec:appdos}, where the corresponding
plots are shown.  The main feature here is that there is a difference
in how varying $z$ affects the DOS of the semiordered systems and of
the ordered systems.  Namely, for semiordered systems
(Figure~\ref{fig:dos} left half), there are significant changes of the DOS, which go
beyond a simple increase of the broadening with increasing $z$ ---
there is a true change of the shape of the DOS curves.  On the other
hand, fully disordered systems (Figure~\ref{fig:dos} right half), the
influence of~$z$ 
on the DOS is more like a gradual broadening of the DOS with
increasing $z$, reflecting the simple fact that large $z$ means more
doping atoms, which leads to a larger substitutional disorder and
therefore large broadening.

\begin{figure*}
\includegraphics[width=17.5cm]{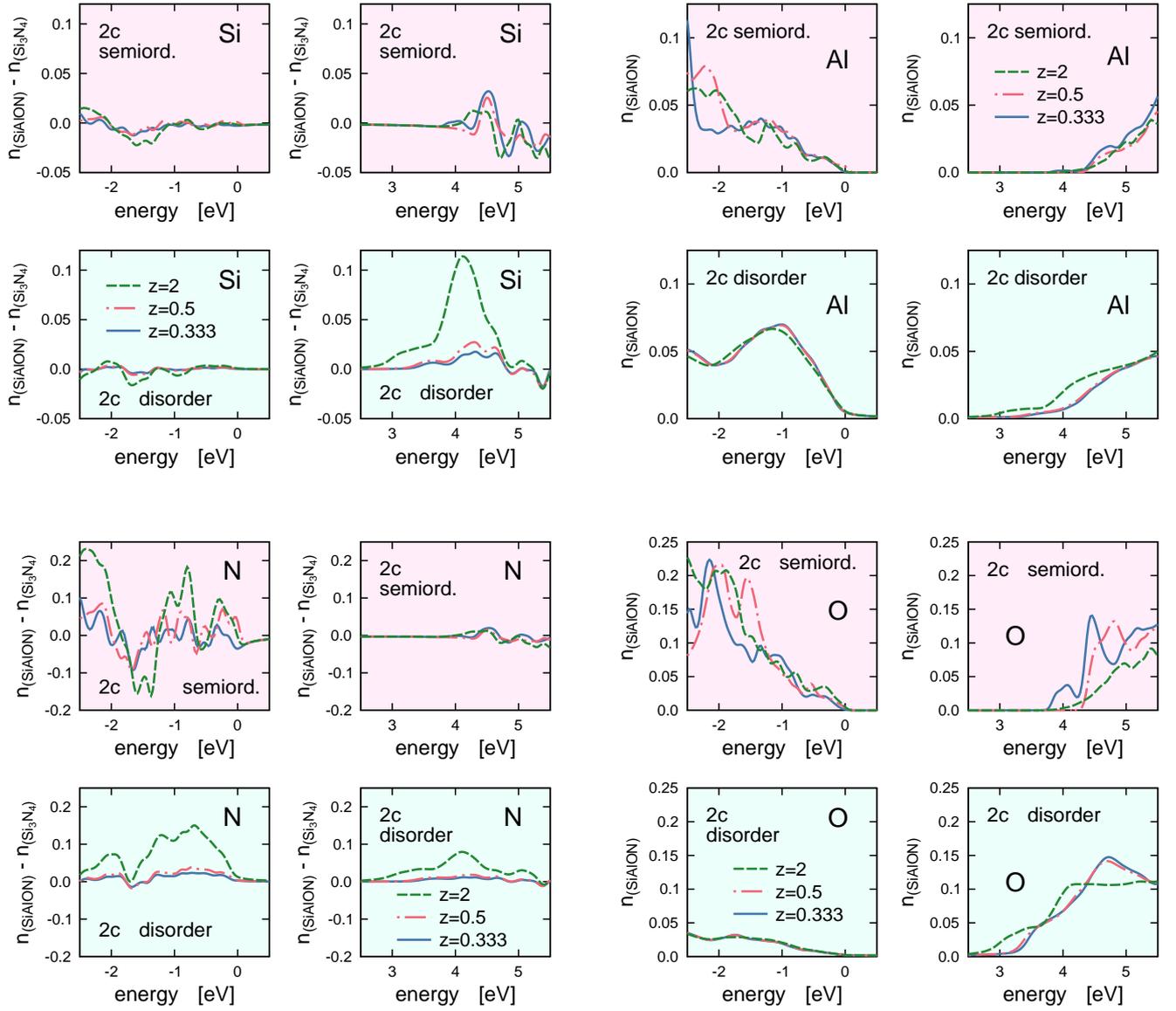}%
\caption{Changes in the DOS induced by adding (Al,O)
  atoms to \bsi.  Changes for Si atoms are shown in the four upper
  left graphs and for N atoms in the four lower left
  graphs. Variations of the DOS for Al atoms are shown in the four
  upper right graphs and for O atoms in the four bottom right graphs.
  DOS for the top of the valence band is explored in the left graphs
  for the respective element and DOS for the bottom of the conduction
  band in the right graphs.  Both semiordered and disordered systems
  are investigated.  For each element, the data were obtained by
  averaging over all sites of the given chemical type in the unit
  cell.  The unit is states per eV.}
\label{fig:diff}
\end{figure*}

Our focus, however, is on how the element-specific DOS varies upon addition of
  (Al,O) atoms.
Therefore we inspect for Si and N atoms the {\em difference} between the DOS for
\bsi\ and for parental \betasin,
\begin{equation}
  n_{\mathrm{Si}_{6-z}\mathrm{Al}_{z}\mathrm{O}_{z}\mathrm{N}_{8-z}}(E) \: - \:
  n_{\mathrm{Si}_{3}\mathrm{N}_{4}}(E)
  \; .
  \label{eq:dif}
\end{equation}
Changes at Si and N atoms are shown in graphs in the left half of
  Figure~\ref{fig:diff}.  For Al and O atoms, the difference
  \mm{n_{\mathrm{Si}_{6-z}\mathrm{Al}_{z}\mathrm{O}_{z}\mathrm{N}_{8-z}}
    - n_{\mathrm{Si}_{3}\mathrm{N}_{4}}}\ coincides with the DOS
  \mm{n_{\mathrm{Si}_{6-z}\mathrm{Al}_{z}\mathrm{O}_{z}\mathrm{N}_{8-z}}}
  because the parental \betasin\ contains no Al or O atoms;
  corresponding data are shown in the right graphs of
  Figure~\ref{fig:diff}.

For each element, the results were obtained by averaging over all
sites of the given chemical type in the unit cell.    We show only the results for
configurations with O atoms at 2$c$ positions in
Figure~\ref{fig:diff}, because the results for
configurations with O atoms at 6$h$ positions exhibit very similar
trends.  Unlike in Figure~\ref{fig:dos}, we do not
weight the DOS by the concentrations of the respective elements in
Figure~\ref{fig:diff}. 

One can see from Figure~\ref{fig:diff} that the way the
  electronic states are influenced by doping the host with (Al,O)
  strongly depends on whether the states are located at Si atoms or at
  N atoms, whether the states are energetically at \tvb\ or at \bcb,
  and whether the system is semiordered or fully disordered.  In
  particular, for Si atoms, occupied states at \tvb\ practically do
  not change if Si$_{3}$N$_{4}$ is doped with (Al,O) whereas
  unocccupied states at \bcb\ show differences upon the doping.  On
  the other hand, states associated with N atoms are more prone to
  changes upon doping at \tvb\ than at \bcb. The differential DOS
  curves \ndiff\ for semiordered systems generally exhibit more fine
  structure than for disordered systems.

For understanding luminescence properties of rare-earths doped \bsi,
it is important to know the character of states adjacent to the gap,
in particular what it the role played by atoms of different chemical
types.
Based on the DOS (Figure~\ref{fig:dos}), we conclude
  that if the multiplicity of sites for each chemical type is taken
into account, the top of the valence band in \bsi\ is dominated by N states and
the bottom of the conduction band is dominated by N and Si states.
This agrees with the results of Hirosaki \ea\ \cite{HKO+05}, Sevik and
Bulutay \cite{SB+07}, or Wang \ea\ \cite{WYC+16}.  On the other hand,
earlier calculations of Xu and Ching \cite{XC+95} provided a bit
different picture; in their study the importance of Si states is
significantly higher than what follows from our data presented in
Figure~\ref{fig:dos}.

Further concerning the character of states at \bcb, our results shown in
  Figure~\ref{fig:diff} demonstrate that if (Al,O) atoms are
  introduced to \betasin, the resulting change in the DOS at Si atoms
  is similar in magnitude to the DOS at Al atoms and, similarly, the
  change in the DOS at N atoms is similar in magnitude to the DOS at N
  atoms.
This indicates that the states at the bottom of the
conduction band are extended, solid-like.  We do not see signatures of
localized impurity-like states reported by Boyko \ea\ \cite{BGS+14} or
Wang \ea\ \cite{WLQ+19}.  Similar conclusions (extended character of
the low-lying unoccupied states) could be drawn also from some other
earlier calculations \cite{HKO+05,SB+07,WYC+16}.

Additional view is obtained by a quantitative assessment of the
relative importance of atoms of given chemical type if the
multiplicity of sites is compensated for.
This is done by integrating the local DOS within a certain energy
  range at \tvb\ and at \bcb, as specified in detail in the
  Appendix~\ref{sec:local}.  Complete results are presented in
  Figure~\ref{fig:locvalcon}.
Here we summarize that states at the top of
the valence band are strongly localized at N atoms, for any $z$, with
no other chemical type giving a significant contribution.  States at
the bottom of the conduction band are mostly localized at O~atoms but
their dominance is not so striking.  If the concentration of
O~atoms~$z$ increases, types other than O are getting also important.
Semiordered and disordered configurations provide the same picture
about attributing states around the gap to chemical types.

Linked to this is the question about the mechanism how the band gap
decreases with increasing~$z$.  Results presented in this section
indicate that this is mostly due to changes at \bcb.


\subsection{Dependence of the DOS on the temperature}   

\label{sec:temp}

\begin{figure}
\includegraphics[width=5.0cm]{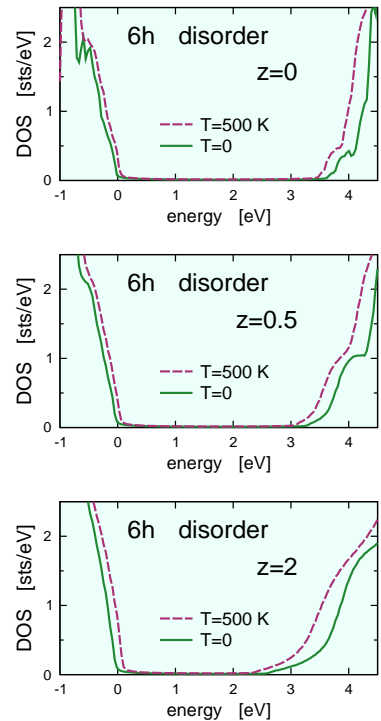}%
\caption{Detailed view on the total DOS of disordered
  \bsi\ with O atoms in $6h$ positions, for $T=0$~K and for $T=500$~K
  around the gap, for different $z$. }
\label{fig:tempgap}
\end{figure}

In this section we explore how the DOS varies under the influence of
temperature.  Only disordered systems are considered because these
systems are investigated using the \kkr\ code for which the alloy
analogy model for dealing with temperature effects has been
implemented.  Based on the results obtained for $T$=0~K (see
Sec.\ref{sec:dos} above), we expect that the
conclusions drawn about the temperature-dependence of DOS on the
temperature for disordered systems can be applied to semiordered
systems as well.

An overall comparison of the DOS for $T$=0~K and for $T$=500~K is
  presented in the Appendix~\ref{sec:apptemp} (Figure~\ref{fig:tempdos}).
Generally, finite temperature broadens the DOS features, as expected.
A more interesting
  aspect is the impact of temperature on the gap.  To highlight it,
we present in Figure~\ref{fig:tempgap} a more detailed view on the DOS
of \bsi\ on a smaller energy scale.  It is evident that the decrease
of $E_{g}$ with increasing $T$ is achieved mainly by changes in the
unoccupied states.  There are no significant differences between the
temperature effects for O atoms at $6h$ positions and at $2c$
positions so we show only results for the first class of systems.

\begin{figure}
\includegraphics[width=8.4cm]{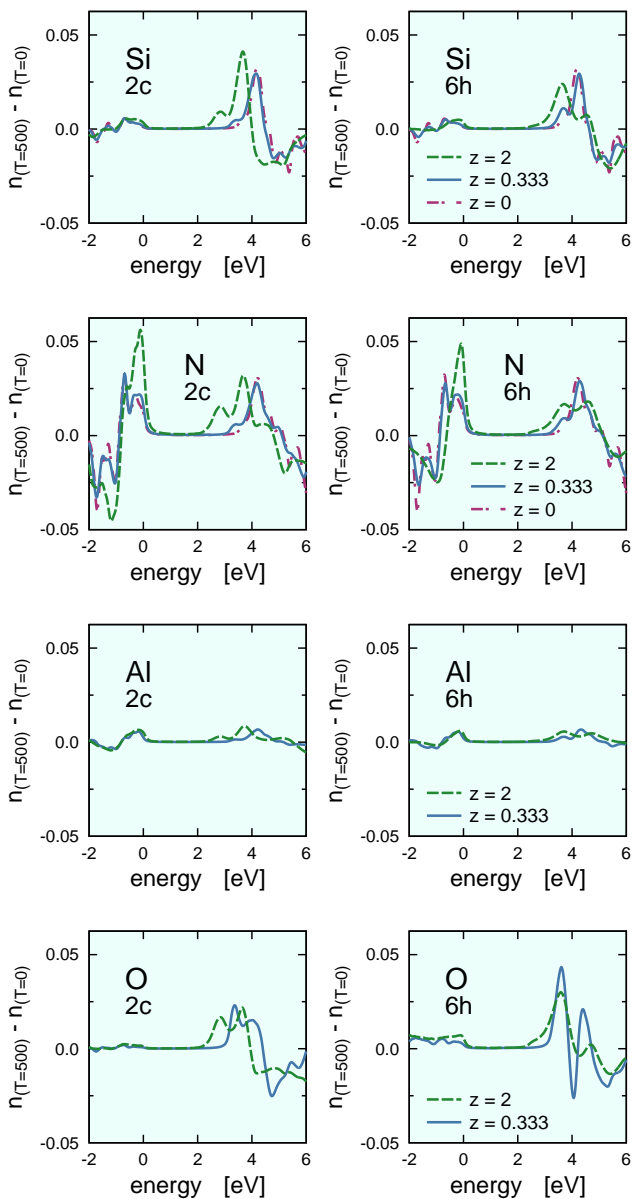}%
\caption{Difference between the local DOS for $T=500$~K
  and for $T=0$~K, $n_{T=500} - n_{T=0}$, for disordered systems with
  O atoms at $2c$ positions. Data for each atomic type were obtained
  by averaging the data for all sites where atoms of the respective
  chemical type occur.  The unit is states per eV.}
\label{fig:tempdif}
\end{figure}

To get an even more focused view, we show in Figure~\ref{fig:tempdif}
the difference between the element-specific DOS calculated
for zero temperature and for $T$=500~K,
\[
n_{T=500}^{(i)}(E) \: - \: n_{T=0}^{(i)}(E)
\; .
\]
Data for parental \betasin\ ($z$=0) and for disordered
\bsi\ with~$z$=0.333 and $z$=2 are shown together, so that one can
monitor how the sensitivity of the DOS to the temperature varies
with~$z$.

One can observe from Figure~\ref{fig:tempdif} that significant
temperature-induced changes for {\em occupied} states occur only at
N~atoms.  Significant temperature-induced changes for {\em unoccupied}
states occur at Si~atoms, N~atoms, and O~atoms.  DOS at Al atoms is
robust with respect to the influence of the temperature.  This is
surprising because Al atoms have similar mass as Si atoms, so there is
no obvious reason why the density of Al-related states should be less
prone to temperature effects than the density of Si-related states.
For O~atoms, there is a visible difference between the effect of
finite temperature on the unoccupied DOS depending on whether they are
at $2c$~positions and whether they are at $6h$~positions.

Another trend seen in Figure~\ref{fig:tempdif} is perhaps more relevant
from the materials science perspective: the amount by which the finite
temperature shifts \bcb\ towards lower energies increases with
increasing~$z$.  In other words, systems with a larger (Al,O)
concentration are more sensitive to the temperature (as concerns the
DOS).  We checked that this trend is valid also for smaller~$z$
(between $z$=0.03 and $z$=0.24, data not shown); there is no reversal
of the trend with~$z$ which would be analogous to the non-monotonous
behavior of the thermal ionization barrier as calculated by Wang \ea\
\cite{WYC+16}.

The results presented here may be helpful for understanding
luminescence quenching phenomena.  A bigger downward shift of
\bcb\ suggests a larger thermal quenching of the luminescence in the
common model of Dorenbos \cite{Dor+05}. Our data suggest that the
thermal quenching should increase with increasing (Al,O)
concentration~$z$.  One should note, however, that this view follows
from analyzing only the host states; in luminescent \bsi-based
materials, the effect of the temperature on the 5$d$ levels of the
rare earth ion and their relative positions to the conduction band
also has to be taken into account.


\section{Conclusions}    \label{sec:concl}

The top of the valence band in \bsi\ is dominated by N states.
The bottom of the conduction band is dominated by N states together
with Si states.  This is --- to a large extent --- because the number
of Si and N atoms is significantly larger than the number of Al
and O atoms, even for the largest~$z$ considered.  If one compensates
for the different numbers of atoms of different types, states at
the top of the valence band are still localized at N atoms.  
States at the bottom of the conduction band are mostly
derived from O~atoms if DOS normalized per site is considered but
their dominance is not overwhelming.
These aspects do not depend on whether the system is only
  semiordered of fully disordered.  Likewise, the influence of whether
  the O atoms are in the $2c$ or in the $6h$ positions is small.

The band gap \eg\ decreases with increasing $z$, by about 1.5~eV when
going from~$z$=0 to~$z$=2.  For semiordered systems it matters 
  whether O atoms are at $2c$ or at $6h$
  positions: in the first case, the decrease of $E_{g}$ with~$z$ is
significantly slower than in the second case.  The importance of
  whether O atoms are in $2c$ or in $6h$ positions decreases  with
  increasing disorder. The scatter of values of 
\eg\ for supercells with the same~$z$ but different distributions of
(Al,O) atoms is quite large --- unlike what is observed for relaxed 
lattice constants.

The decrease of \eg\ with increasing~$z$ is mostly due to changes at
\bcb, which is shifted to lower energies.  Unoccupied states at the
bottom of the conduction band are more affected by presence of Al and
O atoms than occupied states at the top of the valence band.  The
bottom of the conduction band is not impurity-like, it is formed by
states residing on all atomic types.

Increasing the temperature shifts \bcb\ to lower energies.  The amount
of this shift increases with increasing~$z$.  Finite
temperature effects on the DOS are more significant for unoccupied
states at \bcb\ than for occupied states at \tvb.
Systems with larger $z$ are in this regard more sensitive to the
  temperature than systems with low~$z$.

As a whole, we summarize that different local geometries
  corresponding to the same (Al,O) concentration $z$ may lead to quite
  large scatter of electronic structure properties.  Real-life 
  \bsi\ samples will always contain some disorder and hence a
  multitude of local geometries.  It is conceivable that different
  preparation methods may lead to different distributions of (Al,O)
  atoms among the Si and N sites, resulting in different electronic
  properties exhibited by particular samples.  This may contribute
  to sometimes unclear or even contradictory results obtained by
  different studies.


\begin{acknowledgments}
  We would like to thank prof.~W.~Schnick for introducing us to the
  topic and for providing encouragement and inspiration in this line
  of research.  Stimulating discussions were held also with R.~Niklaus
  and P.~J.~Schmidt.  The work was supported by the GA~\v{C}R via the
  project 20-18725S and by the Ministry of Education, Youth and Sport
  (Czech Republic) via the project CEDAMNF
  CZ.02.1.01/0.0/0.0/15\_003/0000358.
\end{acknowledgments}



\appendix



\section{Computational details}    \label{sec:compdets}

The calculations were performed  relying on the generalized gradient
approximation (GGA) using the Perdew, Burke and Ernzerhof (PBE)
functional \cite{Perdew+96}.  When doing geometry optimization by
means of the {\sc 
  vasp} code  \cite{KF+96}, we adopted energy cutoff
700~eV. For the Brillouin zone sampling, the $\Gamma$-centered scheme
in tetrahedron method with Bl\"{o}chl corrections was used, with
3$\times$3$\times$2 $\bm{k}$-points grid for the 1$\times$1$\times$3
supercell ($z$=0.333) and 3$\times$3$\times$3 $\bm{k}$-points grid for
the 1$\times$1$\times$2 supercells ($z$=0.5, 1, and~2).  
The geometries of the supercells representing the semiordered systems
have been relaxed using the conjugate gradient method, requiring
forces less than $10^{-3}$~eV/\AA.

Once the supercells representing the semiordered systems had been
structurally relaxed, their electronic structure was calculated by
means of the FLAPW method implemented in the {\sc wien2k} code
\cite{Blaha+01}.  Wave functions in the interstitial region were
expanded in plane waves, with the plane wave cutoff chosen so that
$R_{\text{MT}}K_{max}$=7.0 ($R_{\text{MT}}$ represents the smallest
atomic sphere radius and $K_{max}$ is the magnitude of the largest
wave vector). The $R_{\text{MT}}$ radii were taken as 1.51~a.u.\
for~Si, 1.51~a.u.\ for~Al, 1.59~a.u.\ for~N, and 1.75~a.u.\ for~O. The
wave-functions inside the spheres were expanded in spherical harmonics
up to the maximum angular momentum
$\ell_{\text{max}}^{\text{(FLAPW)}}$=10.  The $\bm{k}$-space
integration was performed via a modified tetrahedron integration
scheme.  Self-consistent calculations were performed using a grid of
$\bm{k}$-points distributed as 6$\times$6$\times$15 for the standard
\sin\ hexagonal unit cell, 6$\times$6$\times$7 for the \two\
supercell, and 6$\times$6$\times$4 for the \three\ supercell.  When
computing the density of states (DOS), we used a $\bm{k}$-mesh
distributed as 10$\times$10$\times$24 for the \sin\ unit cell,
10$\times$10$\times$11 for the \two\ supercell, and
10$\times$10$\times$7 for the \three\ supercell.

Systems with substitutional disorder were treated by means of the
Green-function multiple-scattering Korringa-Kohn-Rostoker (KKR) method
using the {\sc sprkkr} code \cite{sprkkr-code}.  The disorder was
treated within the so-called coherent potential approximation (CPA):
the system is substituted by an auxiliary effective medium chosen by
requiring that if a real atom is embedded in this medium, no further
scattering is produced on the average \cite{EKM11}. The potential was
subject to the atomic spheres approximation (ASA).  For the multipole
expansion of the Green function, an angular momentum cutoff
\mm{\ell_{\mathrm{max}}^{\text{(KKR)}}}=2 was used.  The
$\bm{k}$-space integration was carried out via sampling on a regular
$\bm{k}$-mesh.  Self-consistent calculations were performed using a
grid of 20$\times$20$\times$20 points in the full Brillouin zone,
density of states was then calculated using a grid of
24$\times$24$\times$24 points.

Finite temperature effects were included by the alloy analogy model
\cite{EMC+15}.  The atomic potentials were considered as frozen.
Atomic vibrations were described using 14 displacement vectors, each
of them being assigned the same probability.  The lengths of these
displacement vectors were set so that, on the average, they reproduce
the temperature-dependent root mean square displacement
\mm{\sqrt{\langle u^{2} \rangle}}\ as given by the Debye theory
\cite{EMC+15}.  For this one needs to know the Debye temperature
$\Theta_{D}$.  Using the calculations of Brgoch \ea\ \cite{BGB+14}, we
take $\Theta_{D}$=954~K for~$z$=0, $\Theta_{D}$=923~K for~$z$=0.333,
$\Theta_{D}$=907~K for~$z$=0.5, $\Theta_{D}$=858~K for~$z$=1, and
$\Theta_{D}$=759~K for~$z$=2.  Displacements for different elements
were set so that differences in their atomic masses are taken into
account.

\begin{figure*}
\includegraphics[width=17.5cm]{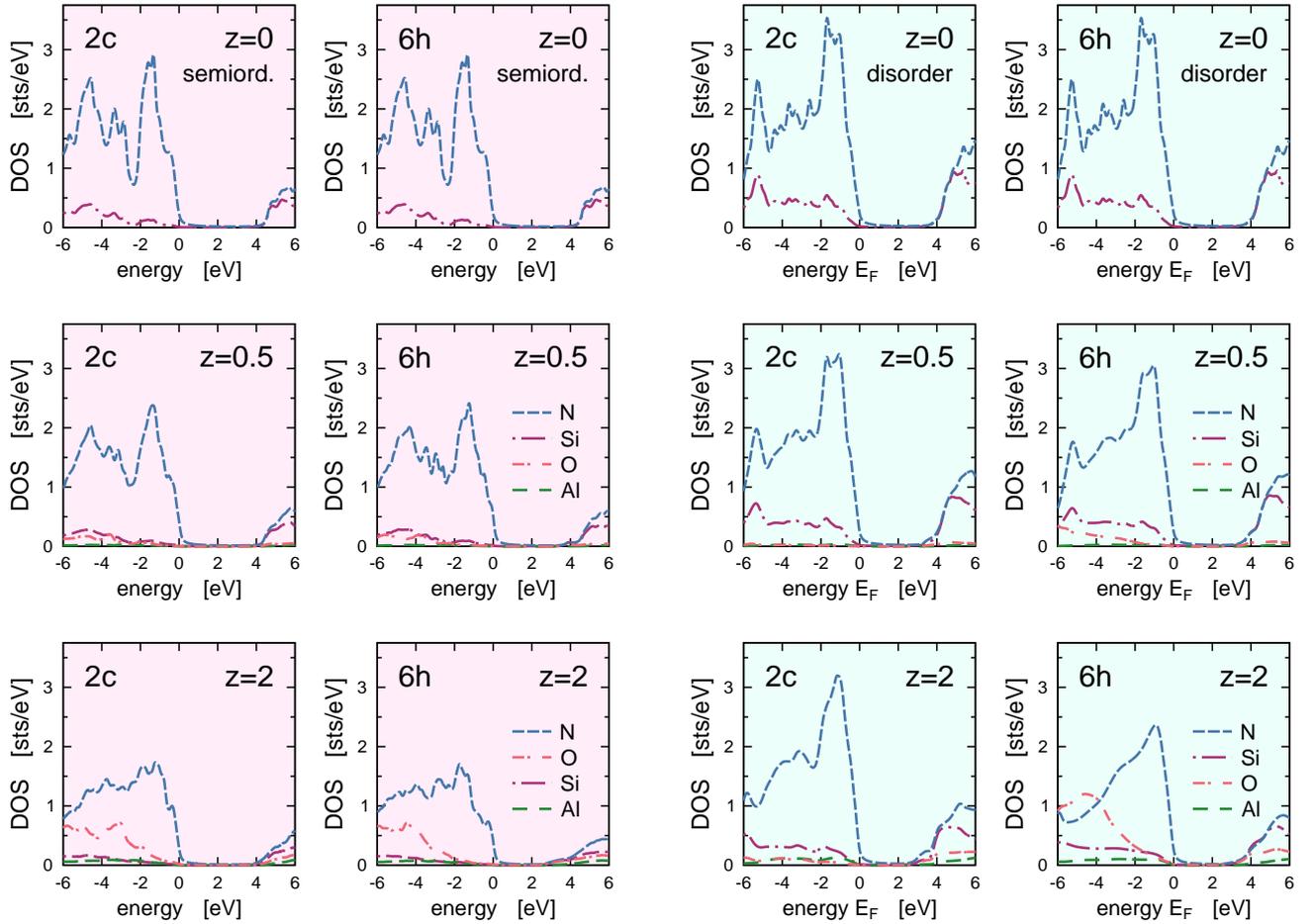}%
\caption{DOS for \bsi\ for $z$=0, $z$=0.5, and $z$=2,
  resolved according to chemical types.  Semiordered systems are
  explored in graphs in the left two columns, disordered systesm are
  explored in graphs in the right two columns.  For semiordered
  systems, each curve represents average over six
  supercells. Multiplicity of sites is taken into account, so DOS for
  Si atoms comprises 6-$z$ sites per unit cell, DOS for Al atoms
  $z$~sites, DOS for N atoms 8-$z$ sites, and DOS for O atoms
  $z$~sites.  Data for configurations with O~atoms at 2$c$ positions
  as well as at 6$h$ positions are shown, as indicated by the
  legends. }
\label{fig:dos}
\end{figure*}

When comparing the DOS for semiordered and disordered systems, one should
keep in mind that the element-specific DOS is defined in a bit
different way in the \wtk\ code and in the \kkr\ code.  Generally,
site-specific DOS $n^{(i)}(E)$ is defined as an integral of the local
DOS,
\begin{equation}
  n^{(i)}(E) \: = \:
  \int_{V_{i}} \! \sum_{n} \, |\psi_{n}(\bm{r})|^{2} \, 
  \delta (E-E_{n})
  \; ,
\label{eq-local}
\end{equation}
within a suitable volume $V_{i}$ around the site $i$.  When using
\wtk, this volume is the appropriate muffin-tin sphere of radius
$R_{\text{MT}}^{(i)}$.  When using \kkr, is it the respective atomic
sphere.  The volume of the atomic sphere employed by \kkr\ is larger
than the volume of the muffin-tin sphere employed by \wtk, which is
reflected also in respective element- or site-specific densities of
states.  This is just a technical issue affecting the absolute values
of the DOS data, it has no implications for the conclusions.


\section{Further results on the dependence of DOS on $z$}

\label{sec:appdos}

An overall view on how the DOS changes upon varying the (Al,O)
concentration~$z$ is presented in Figure~\ref{fig:dos}, both for
semiordered and for disordered systems.  The DOS curves for
semiordered systems were obtained by averaging over all sites of the
respective chemical type and also over the six supercells representing
the structure for each $z$.  We present element-specific DOS, taking
into account the concentrations.  This means that DOS for Si atoms
comprises 6-$z$~sites per unit cell, DOS for Al atoms $z$~sites, DOS
for N atoms 8-$z$~sites, and DOS for O atoms comprises
$z$~sites. Consequently, the dominant contribution to the total DOS
comes from N and Si atoms, because these are most numerous in the unit
cell.  If $z$ increases, the importance of O atoms increases as well
because their number increases.

\begin{figure*}
\includegraphics[width=17.5cm]{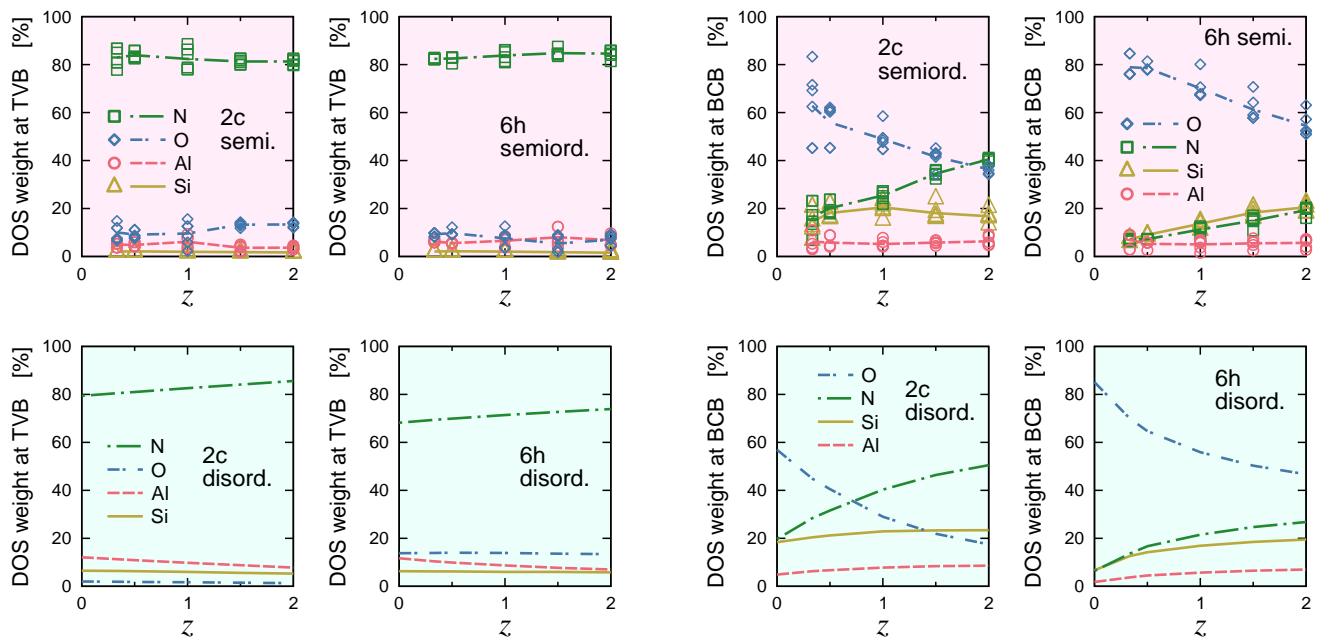}%
\caption{Relative weights of local DOS around the gap
  for atoms of different types, depending on the (Al,O)
  concentration~$z$.  Data for states at the the top of the valence
  band (TVB) are shown in the two leftmost columns, data for states at
  the the bottom of the conduction band (BCB) are shown in the two
  rightmost columns.  Semiordered systems are explored in the upper
  graphs; the markers denote data for the individual supercells
  representing the semiordered system to each $z$, lines denote the
  average values.  Disordered systems are explored in the lower
  graphs.  Left graphs are for configurations with O atoms at 2$c$
  positions, right graphs are for configurations with O atoms at 6$h$
  positions.}
\label{fig:locvalcon}
\end{figure*}

For semiordered systems, there are significant changes of the DOS upon
varying~$z$.  The difference is not just a larger broadening, there is
a true change of the shape of the DOS curves.  For disordered systems,
the influence of~$z$ on the DOS is more like a gradual broadening of
the DOS with increasing $z$, consistently with the notion that large
$z$ means more doping atoms, which leads to a larger substitutional
disorder (see Figure~\ref{fig:dos}).


\section{Quantitative data concerning the localization of states
  around the gap} 

\label{sec:local}

To get study the relative importance of atoms of given chemical type
for states at the top of the valence band and at the bottom of the
conduction band, we evaluate their weights quantitatively.  In
particular the weight $w_{J}$ at the top of the valence band is
evaluated as
\begin{equation}
w_{J} \: = \:
\frac{1}{N_{J}} \, 
\sum_{i=1}^{N_{J}}  \,  
\frac{1}{V_{i}} \, 
\int_{V_{i}} \!\! \dstd \bm{r}  \,
\int_{E_{\text{VB}}-0.4}^{E_{\text{VB}}} \!\! \dstd E \:
n(\bm{r},E)
\; ,
\label{eq:w}
\end{equation}
where $J$ labels the chemical type (Si, N, Al, or O), $N_{J}$ is the
appropriate number of sites in the unit cell, $n(\bm{r},E)$ is the
local density of states whose radial integration covers the volume
$V_{i}$ around the site~$i$ and its energy integration goes from
$E_{\text{VB}}$-0.4~eV to $E_{\text{VB}}$, where $E_{\text{VB}}$ is
the top of the valence band.  In the end, the weights $w_{J}$ are
normalized so that their sum over the chemical types is one.  Weights
at the bottom of the conduction band are evaluated analogously, just
the energy integration goes from $E_{\text{CB}}$ to
$E_{\text{CB}}$+0.4~eV, with $E_{\text{CB}}$ denoting the energy at
the bottom of the conduction band. The {\em ad hoc} chosen energy
integration interval~0.4~eV is not a crucial parameter, we checked
that same conclusions are obtained if it is twice as large or just
half of it.  The factor $1/N_{J}$ in Eq.~(\ref{eq:w}) ensures that we
study here the weights normalized to one atom per unit cell, getting
thus a picture complementary to what is provided by
Figure~\ref{fig:dos} where the multiplicity of
sites in the unit cell was taken into account.

Normalized weights are shown in Figure~\ref{fig:locvalcon} for the top
of the valence band as well as for the bottom of the conduction band.
States at the top of the valence band are clearly localized at N
atoms, for any $z$, with no other chemical type giving a significant
contribution.  States at the bottom of the conduction band are mostly
localized at O~atoms but their dominance is not so striking.  If the
concentration of O~atoms~$z$ increases, types other than O are getting
also important.  This is especially true for configurations with O
atoms at 2$c$ positions, where the N~atoms get more important than
O~atoms for the states at the bottom of the conduction band increase
if $z$ increases.  Semiordered and disordered configurations provide
the same picture about attributing states around the gap to chemical
types.


\vspace*{1.5em}

\section{Dependence of the DOS on the temperature}   

\label{sec:apptemp}

\begin{figure}
\includegraphics[width=8.4cm]{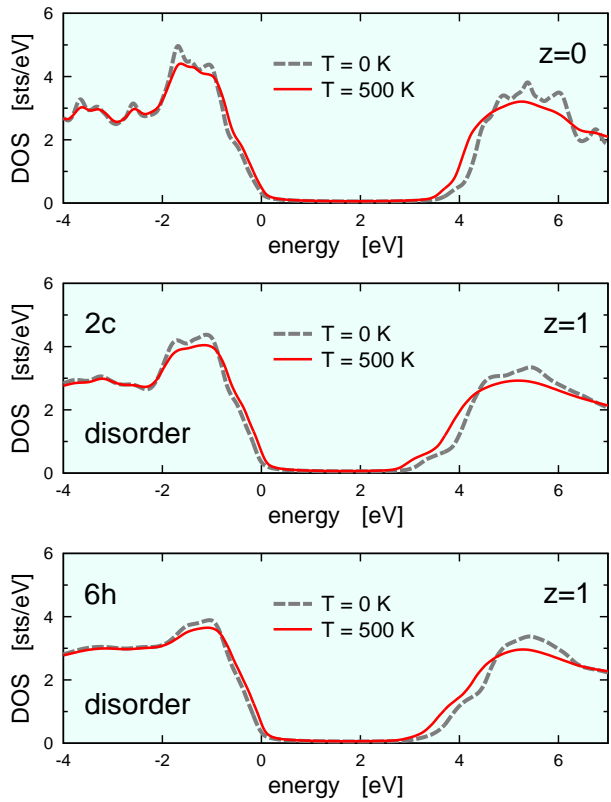}%
\caption{Total DOS of \bsi\ for $T=0$~K and for
  $T=500$~K.  The top graph shows data for $z$=0 (i.e., for \betasin), the
  other two graphs show data for disordered structure for $z$=1, with
  O atoms either at $2c$ positions (the middle panel) or at $6h$
  positions (the bottom panel).  }
\label{fig:tempdos}
\end{figure}

Figure~\ref{fig:tempdos} provides an overall comparison of the DOS for
$T$=0~K and for $T$=500~K.  The effect of temperature on the DOS for
the parental \betasin\ and for disordered \bsi\ with~$z$=1 is similar.
Mostly it consist of smoothening of sharp DOS features.  Additionally,
\bcb\ is shifted to lower energies.  Generally, the influence of
finite temperature on the density of unoccupied states is stronger
than the influence on the density of occupied states.  There are no
significant differences between the temperature effects for O atoms at
$2c$ positions and at $6h$ positions.



\bibliographystyle{apsrev4-1}

\bibliography{liter-sialon}

\end{document}